\documentclass[twoside,twocolumn,10pt]{article}

\usepackage{wscg}           
\RequirePackage{ifpdf}
\ifpdf
 \RequirePackage[pdftex]{graphicx}
 \RequirePackage[pdftex]{color}
\else
 \RequirePackage[dvips,draft]{graphicx}
 \RequirePackage[dvips]{color}
\fi

\usepackage{adjustbox}
\usepackage{amsmath,amssymb}
\usepackage[table,xcdraw]{xcolor}
\usepackage{booktabs}
\usepackage[center]{caption}
\usepackage{multirow}
\usepackage{arydshln}
\usepackage{hyperref}
\hypersetup{breaklinks=true,colorlinks}

\usepackage{nopageno}       

\newcommand{\redb}[1]{\textcolor{red}{\textbf{#1}}}
\newcommand{\ublue}[1]{{\textcolor{blue}{#1}}}

\title{StarSRGAN: Improving Real-World Blind Super-Resolution}

\author{
\parbox{0.4\textwidth}{\centering
Khoa D. Vo\\[1mm]
Faculty of Information Technology (FIT)\\
University of Science, VNU.HCM\\
Ho Chi Minh City, Vietnam\\[1mm]
20c11008@student.hcmus.edu.vn
}
\hspace{0.05\textwidth}
\parbox{0.4\textwidth}{\centering
Len T. Bui\\[1mm]
Faculty of Information Technology (FIT)\\
University of Science, VNU.HCM\\
Ho Chi Minh City, Vietnam\\[1mm]
btlen@fit.hcmus.edu.vn
}
}

\usepackage{url}
\makeatletter
\def\Uslash{\mathbin{\mathchar`\/}\@ifnextchar{/}{\kern-.15em}{}}
\g@addto@macro\UrlSpecials{\do \/ {\Uslash}}
\def\Ucolon{\mathbin{\mathchar`:}\@ifnextchar{/}{\kern-.1em}{}}
\g@addto@macro\UrlSpecials{\do : {\Ucolon}}
\makeatother

\begin{document}

\twocolumn[{
\csname @twocolumnfalse\endcsname

\maketitle  

\begin{center}
    \centering
    \vspace{-0.2cm}
    \includegraphics[width=\linewidth]{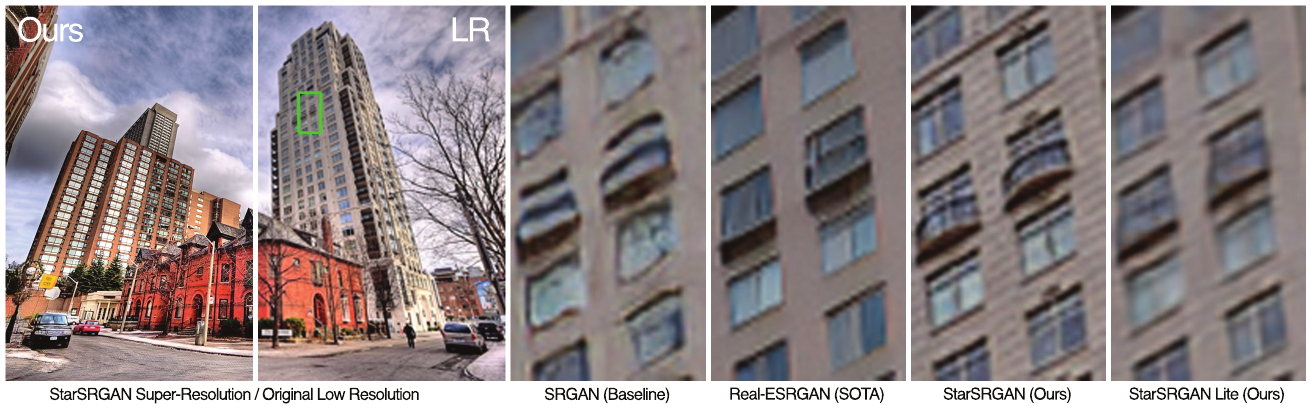}

    \captionsetup{hypcap=false}
    \captionof{figure}{Comparison between our models and some standard SRGAN models. (\textbf{Zoom in for best view})}
    \label{fig:teaser}
\end{center}
\vspace*{1.0\baselineskip}
\begin{abstract}
The aim of blind super-resolution (SR) in computer vision is to improve the resolution of an image without prior knowledge of the degradation process that caused the image to be low-resolution. The State of the Art (SOTA) model Real-ESRGAN has advanced perceptual loss and produced visually compelling outcomes using more complex degradation models to simulate real-world degradations. However, there is still room to improve the super-resolved quality of Real-ESRGAN by implementing recent techniques. This research paper introduces StarSRGAN, a novel GAN model designed for blind super-resolution tasks that utilize 5 various architectures. Our model provides new SOTA performance with roughly 10\% better on the MANIQA and AHIQ measures, as demonstrated by experimental comparisons with Real-ESRGAN. In addition, as a compact version, StarSRGAN Lite provides approximately 7.5 times faster reconstruction speed (real-time upsampling from 540p to 4K) but can still keep nearly 90\% of image quality, thereby facilitating the development of a real-time SR experience for future research.
Our codes are released at \normalsize\url{https://github.com/kynthesis/StarSRGAN}.
\end{abstract}

\subsection*{Keywords}
Blind super-resolution, adaptive degradation, dual perceptual loss, multi-scale discriminator, dropout degradation
\vspace*{1.0\baselineskip}
}]

\section{Introduction}

\copyrightspace

The field of image super-resolution (SR) aims to reconstruct a high-resolution (HR) image from a low-resolution (LR) counterpart, which is a challenging task due to the many-to-one mapping involved. The computer vision research community has devoted significant attention to SR, and deep learning algorithms have been successfully applied to this problem. These techniques utilize neural networks to train an end-to-end mapping function, such as the SRCNN \cite{Don16}, which involves deep convolutional neural networks (CNNs) that generate high signal-to-noise ratio (PSNR) values. Still, the output is often excessively smoothed and needs more high-frequency features.

Researchers suggest using generative adversarial networks (GANs) \cite{Goo14} for image SR tasks to overcome these limitations. A super-resolution GAN comprises a generator network and a discriminator network. The generator takes LR images as input and aims to create images that resemble the original HR image. At the same time, the discriminator distinguishes between "fake" and "real" HR images. However, these methods assume an ideal bicubic downsampling kernel, making them unsuitable for real-world situations.

In contrast, blind SR aims to recover images with unknown and complicated degradations. Existing techniques can be classified as explicit or implicit modelling based on the underlying degradation process. Detailed modelling approaches frequently utilize the conventional degradation model, including blur, downsampling, noise, and JPEG compression. However, real-world degradations need to be simplified to be described by a simple combination of these factors. Current research has focused on emulating a more practical degradation process \cite{Wan21} or designing an improved generator \cite{Wan19}. The performance of the discriminator, which guides the generator to generate superior images, has received relatively little attention. Therefore, it is essential to acknowledge the importance of the discriminator, much like a loss function.

In this paper, we intend to further enhance the perceptual quality of super-resolved images by extending the robust Real-ESRGAN \cite{Wan21} algorithm:

\begin{itemize}
\item We utilize the Star Residual-in-Residual Dense  Block (StarRRDB), which was inspired by ESRGAN+ \cite{Rak20} and had a higher capacity than the RRDB employed by Real-ESRGAN \cite{Wan21}.
\item We obtain a performance breakthrough for our SR models by combining the Multi-scale Attention U-Net Discriminator with the present StarRRDB-based generator inspired by A-ESRGAN \cite{Wei21}.
\item We replaced the standard high-order Real-ESRGAN degradation model with a DASR-inspired \cite{Lia22} efficient Adaptive Degradation Model.
\item We use ResNet Loss in addition to VGG Loss, and this Dual Perceptual Loss approach, inspired by ESRGAN-DP \cite{Son22}, acquires more sophisticated perceptual characteristics.
\item We attempt to apply Dropout Degradation technique, inspired by RDSR \cite{Kon21} to improve the generalization ability from appropriate usage of dropout benefits.
\item We construct StarSRGAN Lite, a CNN-oriented compact version that can reconstruct images about 7.5 times faster than StarSRGAN and Real-ESRGAN.
\end{itemize}

Due to several modifications, our StarSRGAN achieves higher visual performance than Real-ESRGAN, making it more applicable to real-world applications.

\section{Related Work}

\subsection{Super-Resolution Methods}

\begin{figure*}[htp]
\centerline{\adjustbox{max width=\textwidth}{}}
\centerline{
    \adjustbox{max width=\textwidth}{
        \includegraphics[height=\textheight]{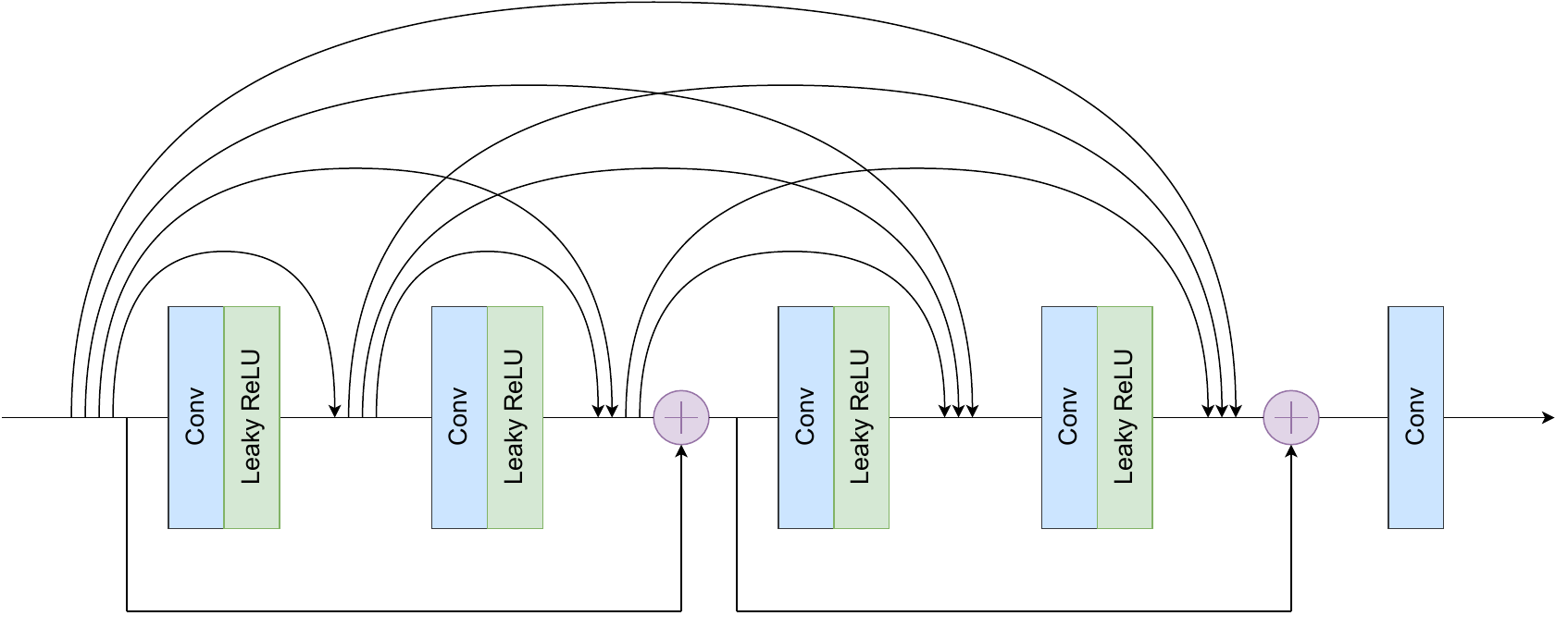}
    }
}
\caption{In Star Dense Block, residuals are added every two layers.}
\label{fig:star-dense}
\end{figure*}

In blind image SR problems, deep CNNs are frequently used before implementing the GAN architecture. These techniques have achieved a remarkable peak signal-to-noise ratio (PSNR) due to the robust modelling capability of CNNs. However, since these PSNR-based approaches use pixel-wise specified losses such as mean squared error (MSE), the output is often overly smoothed, necessitating additional high-frequency information. In practice, most methods assume a bicubic downsampling kernel, which may not be effective for real-world images. Furthermore, current studies aim to incorporate reinforcement learning or GAN before image restoration.

Blind SR has been widely researched, with numerous studies focusing on degradation prediction and conditional restoration. The two processes can be performed separately or in tandem, often iteratively. These techniques rely on predefined degradation models, which may only consider synthetic degradations and fail to perform well with real-world images. Furthermore, inaccurate degradation models can result in unwanted artifacts in the reconstructed images.

Recently, SOTA research has proposed a perceptually-driven approach to improve GANs by more accurately simulating the perceptual loss between images. For instance, ESRGAN \cite{Wan19} and ESRGAN+ \cite{Rak20} have introduced a viable perceptual loss function and generator networks based on RRDB that can convincingly create HR images. Another method, Real-ESRGAN \cite{Wan21}, has introduced a high-order degradation model to make even more realistic images, achieving impressive results on the NIQE benchmark. However, these methods depend on a computationally complex backbone network and cannot handle images with varying levels of degradation. Therefore, DASR \cite{Lia22} has introduced a degradation-adaptive framework to address this issue, creating an effective and efficient network for real-world SR challenges.

Our work has incorporated several benefits from various designs to produce a comprehensive solution.

\subsection{Degradation Models}

Blind SR approaches often rely on the classical degradation model, which may not fully represent the complex degradation in real-world images. Recent technique, such as Real-ESRGAN \cite{Wan21}, incorporate a broader range of degradation types and parameters into the modelling process to address this issue. These approaches increase the model's ability to enhance the perceptual quality of challenging LR inputs. However, the sampling of degradation parameters in these methods can be imbalanced, which limits their ability to generate fine features, particularly for inputs with average degradations.

This study implements an adaptive degradation model that balances the degradation space by dividing it into three frequency-balanced levels. This balanced space optimises the model at different levels and provides a more accurate representation of real-world images.

\subsection{Discriminator Models}

There have been significant efforts to improve the discriminator model to synthesize high-quality HR images. Two critical challenges in achieving photo-realistic HR images are the need for a broad receptive field for the discriminator to distinguish between the synthesized and ground truth images, which requires either a deep neural network or a big convolution kernel. Additionally, a single discriminator may struggle to provide input on global and local characteristics, leading to potential incoherence in the synthesized image, such as distorted wall textures.

pix2pixHD \cite{Wan17} proposed a unique multiple discriminator architecture to address these challenges. The first discriminator takes downsampled synthetic images as input, has a broader receptive field with fewer parameters, and focuses on comprehending the global perspective. The second discriminator uses the entire synthesized image to learn the image's specifics. Another study \cite{Sch20} employed a U-Net-based discriminator architecture to GAN-based blind SR challenges, preserving the global coherence of synthesized images while offering per-pixel input to the generator.

Our discriminator model combines the advantages of both designs, enabling the discriminator to learn edge representations, enhance training stability, and provide per-pixel feedback to the generator.

\subsection{Image Quality Assessment Methods}

In recent years, GAN has been widely utilized for restoring low-quality images (e.g., deblur, denoise, super-resolution). Some researchers have focused on assessing images using Image Quality Assessment (IQA) methods. Some synthetic textures look natural due to the GAN approach, making them difficult for humans to distinguish yet easy for machines to detect.

Attention-based Hybrid Image Quality Assessment Network (AHIQ) \cite{Lao22} aims to quantify the human perception of image quality and has the potential for generalizing unknown and complex samples, notably GAN-based distortions. AHIQ won first place in Full-Reference (FR) Track for the NTIRE2022 Perceptual Image Quality Assessment Challenge \cite{Nti22}.

Multi-dimension Attention Network for No-Reference Image Quality Assessment (MANIQA) \cite{Yan22} uses the multi-dimensional attention network for perceptual assessment. MANIQA placed first in the No-Reference (NR) Track of the NTIRE2022 Perceptual Image Quality Assessment Challenge \cite{Nti22}.

Since no real-world GT exists for SR images, NR-IQA is preferable to FR-IQA for SR visual comparison. Nonetheless, in this study, we use both AHIQ and MANIQA, as well as some classical metrics like PSNR, SSIM, NIQE, and LPIPS. The objective is to determine if StarSRGAN models can achieve new performance levels in many measures.

\subsection{Perceptual Loss Methods}

Since the groundbreaking SRCNN \cite{Don16} was presented, applying deep learning to tackle the SR problem has garnered increasing interest. In addition to a significant boost in visual quality, it also features a greater variety of optimizations and enhancements.

Further research \cite{Joh16} claimed that smoothing the reconstructed image was as simple as improving the MSE or PSNR of the pixel space ratio between the GT image and the reconstructed image. In order to enhance the reconstruction effect, the perceptual loss was suggested to minimize the feature space error between the GT image and the rebuilt image.

Based on the concept of SRGAN \cite{Led17}, ESRGAN \cite{Wan19} employed a range of strategies to enhance the texture features further. In terms of perceptual loss, they recommended using the output before the activation of the convolutional layer to gain additional feature information, with the error of the feature space before activation being the object to be reduced.

Discussing perceptual loss is essential for enhancing the reconstruction outcomes, particularly the realism of details. This work uses a unique dual perceptual loss function as a combination of ResNet \cite{Kai16} loss and VGG \cite{Sim15} loss to achieve the reduction of unnatural artifacts produced by the perceptual-driven technique.

\section{Proposed Methods}

\subsection{Adaptive Degradation Model}

Recently, a high-order degradation model has been proposed to generate LR images more closely approximate real-world conditions. The model executes the same degradation operation multiple times and has advanced from simple bicubic down-sampling to include shuffling and second-order pipelines. 

In this research, we incorporate numerous image degradation procedures, including blurring (both isotropic and anisotropic Gaussian blur), resizing (both down-sampling and up-sampling with area, bilinear, and bicubic operations), noise corruption (both additive Gaussian and Poisson noise), and JPEG compression.

Inspired by the DASR \cite{Lia22} approach, our architecture is designed to be adaptive to a wide range of real-world inputs and handle a subspace of images with different degradation levels. We divide the entire degradation space $D$ into three levels: $[D_1, D_2, D_3]$, with one of these randomly chosen to produce LR-HR image pairs during training. The probability distribution for selecting the levels is [0.3, 0.3, 0.4]. $D_1$ and $D_2$ use first-order degradation with small and large parameter ranges, while $D_3$ uses second-order degradation. We use isotropic and anisotropic Gaussian kernels for the blur operation with a probability of [0.65, 0.35]. If an isotropic blur kernel is supplied, we set $\sigma_1 = \sigma_2$. In the second degradation stage of $D_3$, we skip the blur operation with a 20\% probability and use sinc kernel filtering with an 80\% probability, following the approach used in Real-ESRGAN. We scale the image to the appropriate LR size, a quarter of its original size.

\subsection{Network Architecture}

\textbf{StarSRGAN.} The core block of ESRGAN enables the network to be very scalable and more straightforward to train. The Star Dense Block we suggested is replacing the Dense Block to increase the network's capacity. Figure \ref{fig:star-dense} depicts an extra level of residual learning within the Dense Blocks to expand the capacity without increasing complexity. After two layers, a residual is added to each block. This novel architecture produces images with improved perceptual quality by utilizing feature exploitation and exploration.

\begin{figure}[h]
\centerline{
    \adjustbox{max width=\columnwidth}{
        \includegraphics[height=\textheight]{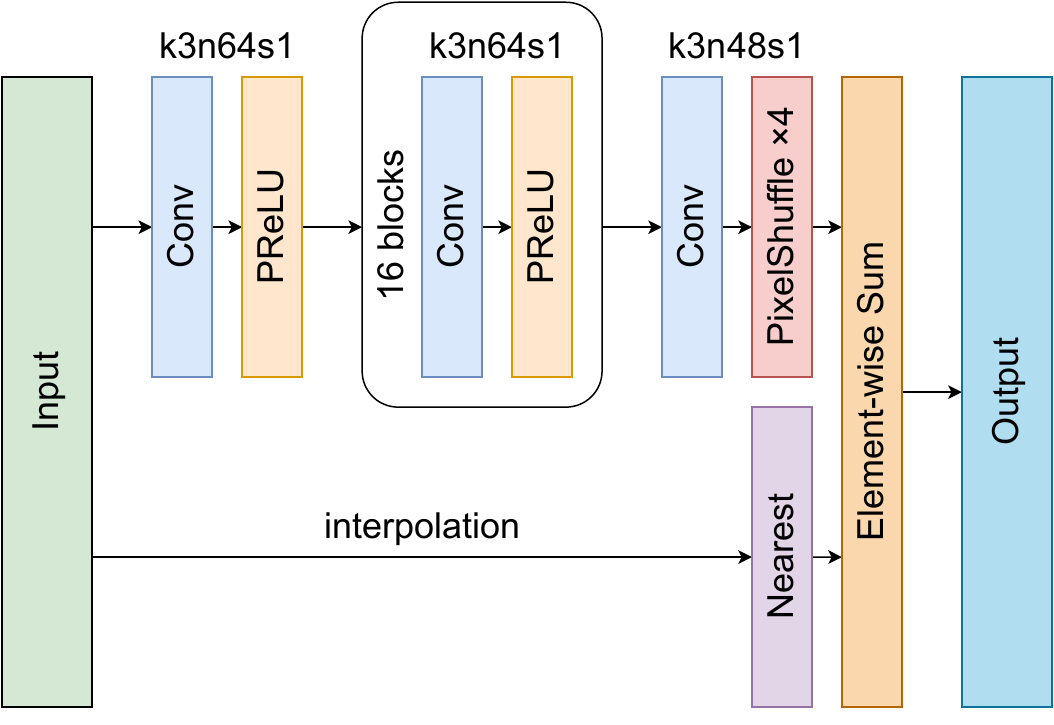}
    }
}
\caption{Architecture of StarSRGAN Lite with corresponding kernel size (k), number of filters (n), and stride (s) indicated for each convolutional layer.}
\label{fig:starsrgan-lite}
\end{figure}

\textbf{StarSRGAN Lite.} The lightweight version of StarSRGAN focuses on delivering faster reconstruction times while maintaining acceptable visual quality. Like its predecessor, the model seeks to capitalize on the Multi-scale Discriminator, Attention U-Net Discriminator, Dual Perceptual Loss, Dropout Degradation, and Adaptive Degradation. ESPCN \cite{Shi16} influences the network design, a super-resolution CNN which brings asymptotic real-time performance. Figure \ref{fig:starsrgan-lite} depicts the architecture of StarSRGAN Lite.

\subsection{Attention U-Net Discriminator}

\begin{figure*}[htp!]
\centerline{
    \adjustbox{max width=\textwidth}{
        \includegraphics[height=\textheight]{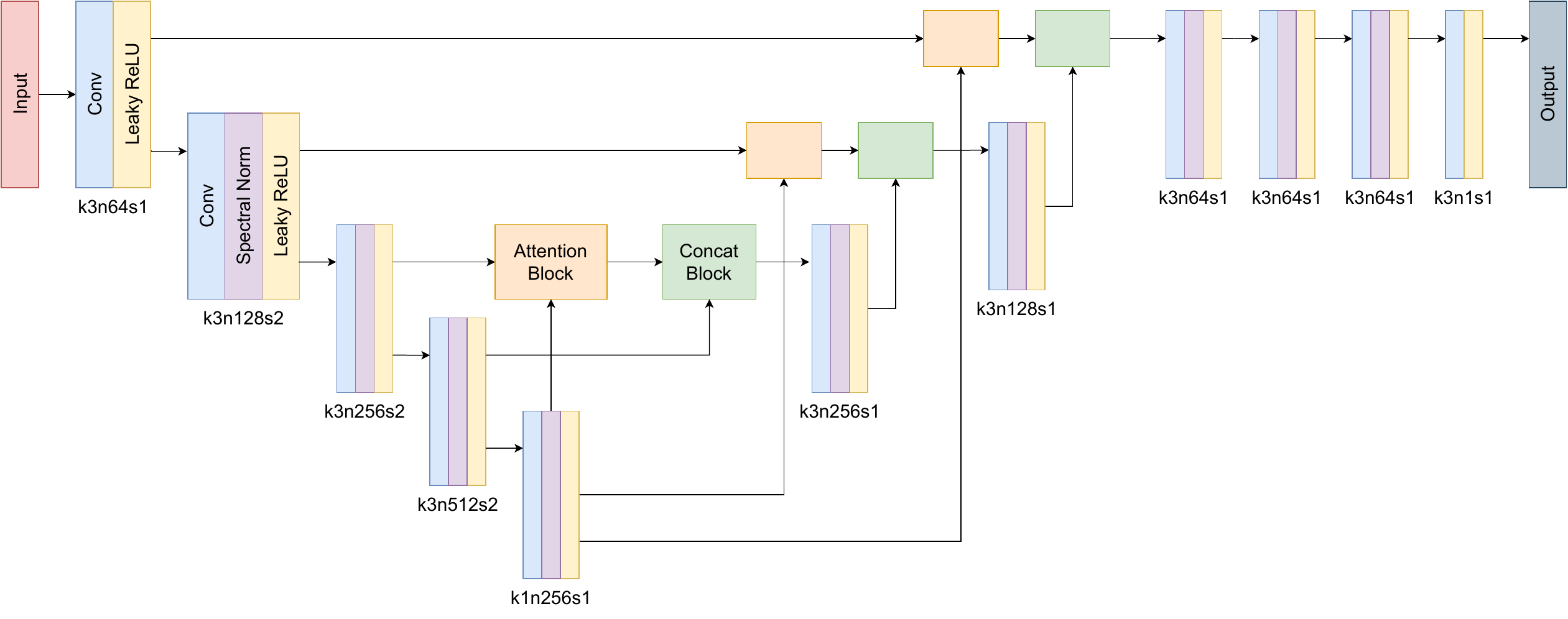}
    }
}
\caption{Architecture of Attention U-Net Discriminator with corresponding kernel size (k), number of feature maps (n), and stride (s) indicated for each convolutional layer.}
\label{fig:attention-unet}
\end{figure*}

Taking inspiration from A-ESRGAN \cite{Wei21}, we have developed an Attention U-Net Discriminator structure, depicted in Figure \ref{fig:attention-unet}, that aims to enhance the quality of the reconstructed image while increasing the efficiency of image reconstruction. The structure comprises a downsampling encoding module, an upsampling decoding module, and multiple Attention Blocks. The Attention and Concatenation Blocks are designed following the A-ESRGAN architecture. To perform semantic segmentation of 2D images, we adapted the Attention Gate, initially scheduled for 3D medical images as described in \cite{Okt18}. Furthermore, we incorporated Spectral Normalization regularization \cite{Miy18} to stabilize the training process.

\subsection{Multi-scale Discriminator}
StarSRGAN has a Multi-scale Discriminator architecture consisting of two identical U-Net discriminators. The first discriminator $D_1$ accepts an original scale image as input, and the second discriminator $D_2$ accepts a $2\times$ downsampled image as input.

The output of the U-Net discriminator is a $W \times H$ matrix, with each member representing the probability that the pixel it represents is True. We utilize the Sigmoid function to normalize the output and the binary cross-entropy loss to determine the overall loss of one discriminator. Assuming $C$ is the output matrix, we define $D=\sigma(C)$, $x_r$ is 'real' data, and $x_f$ is 'fake' data.

Consequently, we define the loss of one discriminator as
\begin{equation}
\begin{split}
    L_D= &\sum_{w=1}^{W}\sum_{h=1}^{H}(-E_{x_r}[\log(D_(x_r,x_f)[w,h])] \\
    &-E_{x_f}[1-\log(D(x_f,x_r)[w,h])])\\    
\end{split}
\end{equation}

Because we have multi-scale discriminators, we will sum the loss of all the discriminators to get the overall loss as
\begin{equation}
    L_{Total}=\lambda_1 L_{D_{normal}}+\lambda_2 L_{D_{sampled}}
\end{equation}
where $\lambda_1$ and $\lambda_2$ are coefficients. We can also derive the generator loss from a single discriminator as
\begin{equation}
\begin{split}
    L_G=&\sum_{w=1}^{W}\sum_{h=1}^{H}(-E_{x_r}[1-\log(D_(x_r,x_f)[w,h])]\\
    &-E_{x_f}[\log(D(x_f,x_r)[w,h])]\\
\end{split}
\end{equation}
where $x_f$ represents the output of the generator $G(x_i)$.

\subsection{Dual Perceptual Loss}

By training a deep neural network, we address the SR problem. According to the theory given by Dong et al. \cite{Don16}, the following is the goal of optimization:
\begin{equation}
\min _{\theta} \frac{1}{n} \sum_{i=1}^{n} L\left(G\left(I_{i}^{L R} ; \theta\right), I_{i}^{H R}\right),
\end{equation}
Where $I_{i}^{L R} \in \mathbf{R}^{C \times H \times W}$ and  $I_{i}^{H R} \in \mathbf{R}^{C \times H \times W}$ represent the $i\mbox{-}th$ LR and HR sub-image pairings, respectively, in the training set. $G\left(I_{i}^{L R} ; \theta\right)$ is the representation of the upsampling network. $\theta$ is the parameter to be optimized within the neural network. $L$ is the loss function that can be represented as:
\begin{equation}
L=\lambda l_{{content }}+\eta l_{{adversarial }}+\gamma l_{percep}
\label{total_loss}
\end{equation}
where $l_{{content}}$ is the content loss of the pixel-wise 1-norm distance between images reconstructed by the generator and GT images, $l_{{adversarial}}$  is the loss derived from the mentioned Multi-scale Discriminator, $l_{{percep}}$  is the Dual Perceptual Loss we implement. $\lambda$, $\eta$, and $\gamma$ are the coefficients of balancing different loss terms.

SRGAN \cite{Led17} proposed defining the VGG loss based on the ReLU activation layer of the pre-trained VGG-19 network. ESRGAN \cite{Wan19} redefined the VGG loss after the convolutional layer and before the activation layer to gain additional feature information. This study uses the VGG loss specified in \cite{Wan19}, as the L1 Norm function is utilized to determine the Manhattan Distance between the reconstructed image features and the GT image features:

\begin{table*}[t]
    \caption{Quantitative comparison with SOTA methods on common test datasets using standard IQAs for SR ($4\times$). LPIPS/NIQE $\downarrow$: the lower, the better. PSNR/SSIM/MANIQA/AHIQ $\uparrow$: the higher, the better. Note that MANIQA and AHIQ are the current SOTA IQA methods. The best and second performance are marked in \redb{red} and \ublue{blue}.\\
    \textbf{Discussion}: The unusual city textures on Urban100 are generally intricate and should consider further fine-tuning for particular usage. Models using real-world emulated data synthesis failed on Manga109 (Japanese comic) dataset, as prior predictions. Thanks to the Adaptive Degradation Model have helped StarSRGAN solve this problem partially. StarSRGAN has a slightly lower performance on Set14, which contains many drawing pictures.} \label{tab:synthetic}
    \setlength{\tabcolsep}{2pt}
    \renewcommand*{\arraystretch}{1.2}
    \resizebox{\textwidth}{!}{
    \begin{tabular}{|c|cccc|cc|cccc|cc|cccc|cc|}
    \hline
    \multirow{2}{*}{Method} & \multicolumn{6}{c|}{DIV2K} & \multicolumn{6}{c|}{Set5} & \multicolumn{6}{c|}{Set14} \\ \cline{2-19}
     & PSNR & SSIM & LPIPS & NIQE & MANIQA & AHIQ & PSNR & SSIM & LPIPS & NIQE & MANIQA & AHIQ & PSNR & SSIM & LPIPS & NIQE & MANIQA & AHIQ \\ \hline
    Original HR & 80.00 & 1.0000 & 0.0000 & 3.2342 & 0.6542 & 0.6289 & 80.00 & 1.0000 & 0.0000 & 3.2122 & 0.6338 & 0.6353 & 80.00 & 1.0000 & 0.0000 & 5.7318 & 0.5807 & 0.6402 \\ \hdashline
    SRGAN & \redb{25.71} & 0.7076 & 0.1808 & \ublue{3.0079} & 0.5256 & 0.4847 & \redb{30.64} & 0.8361 & \ublue{0.1080} & 4.3067 & 0.5084 & 0.5359 & \redb{19.94} & 0.4688 & 0.2566 & \ublue{2.5155} & 0.5106 & 0.4865 \\ 
    ESRGAN & 24.94 & 0.6827 & \ublue{0.1471} & 3.1900 & 0.5616 & 0.4938 & \ublue{29.92} & 0.8258 & 0.1117 & 3.9433 & 0.5396 & 0.5461 & 18.60 & 0.4209 & \ublue{0.2039} & \redb{2.3251} & 0.5329 & 0.5203 \\
    Real-ESRGAN & 24.23 & 0.6646 & 0.2637 & 3.1461 & \ublue{0.5728} & \ublue{0.5112} & 26.06 & 0.7806 & 0.1901 & 3.6883 & 0.5794 & 0.5524 & 18.00 & 0.4186 & 0.3578 & 3.6056 & \ublue{0.5979} & 0.5233 \\ \hdashline
    A-ESRGAN & 23.12 & 0.6032 & 0.3110 & \redb{2.9722} & 0.5398 & 0.4217 & 23.95 & 0.6980 & 0.2060 & \redb{2.8405} & 0.6046 & 0.4730 & 18.12 & 0.3476 & 0.4927 & 2.7213 & \redb{0.6017} & 0.5245 \\  
    FeMaSR & 22.10 & 0.6178 & 0.2236 & 4.9289 & 0.5383 & 0.4426 & 24.22 & 0.7202 & 0.1943 & 4.4306 & \ublue{0.6101} & 0.5054 & 17.49 & 0.3872 & 0.3314 & 3.6338 & 0.5884 & 0.4968 \\
    Swin2SR & \ublue{25.65} & \ublue{0.7164} & 0.4282 & 5.8883 & 0.4346 & 0.4319 & 28.53 & \redb{0.8629} & 0.2855 & 6.9146 & 0.5123 & \ublue{0.5681} & \ublue{19.94} & \ublue{0.4756} & 0.2556 & 2.6782 & 0.5689 & \ublue{0.5307} \\ \hdashline
    \textbf{StarSRGAN} & 25.53 & \redb{0.7232} & \redb{0.1365} & 3.2513 & \redb{0.6151} & \redb{0.5529} & 29.60 & \ublue{0.8576} & \redb{0.0905} & \ublue{3.3224} & \redb{0.6195} & \redb{0.6160} & 19.22 & \redb{0.5167} & \redb{0.2035} & 3.2480 & 0.5932 & \redb{0.5661} \\
    \textbf{StarSRGAN Lite} & 24.36 & 0.6708 & 0.2588 & 4.3514 & 0.5016 & 0.4128 & 28.19 & 0.7995 & 0.1731 & 3.7280 & 0.5249 & 0.5309 & 17.91 & 0.4572 & 0.2780 & 4.6240 & 0.5284 & 0.5011 \\ \hline \hline
    \multirow{2}{*}{Method} & \multicolumn{6}{c|}{BSD100} & \multicolumn{6}{c|}{Urban100} & \multicolumn{6}{c|}{Manga109} \\ \cline{2-19}
     & PSNR & SSIM & LPIPS & NIQE & MANIQA & AHIQ & PSNR & SSIM & LPIPS & NIQE & MANIQA & AHIQ & PSNR & SSIM & LPIPS & NIQE & MANIQA & AHIQ \\ \hline
    Original HR & 80.00 & 1.0000 & 0.0000 & 3.3586 & 0.8206 & 0.6761 & 80.00 & 1.0000 & 0.0000 & 1.8932 & 0.7743 & 0.6395 & 80.00 & 1.0000 & 0.0000 & 3.0512 & 0.7282 & 0.7055 \\ \hdashline
    SRGAN & \ublue{20.39} & 0.6455 & 0.2679 & 2.8920 & 0.6461 & 0.3503 & 18.98 & 0.6557 & 0.1617 & \redb{1.6539} & 0.6303 & 0.49381 & 22.58 & 0.6822 & \ublue{0.1398} & \redb{2.4046} & 0.5991 & 0.5039 \\ 
    ESRGAN & 20.26 & 0.6187 & \ublue{0.2391} & \redb{2.5170} & 0.6684 & 0.3795 & 18.06 & 0.6223 & \ublue{0.1435} & 2.4639 & 0.6753 & \ublue{0.5204} & \ublue{22.70} & 0.6735 & \redb{0.1239} & \ublue{2.4314} & 0.6266 & \ublue{0.5252} \\ 
    Real-ESRGAN & 19.35 & 0.6127 & 0.3183 & 4.6826 & \ublue{0.6815} & \ublue{0.4029} & 17.75 & 0.5966 & 0.2205 & 4.9107 & \ublue{0.7068} & 0.4643 & 20.26 & 0.6262 & 0.2203 & 4.6752 & \ublue{0.6883} & 0.4366 \\ \hdashline 
    A-ESRGAN & 18.52 & 0.5729 & 0.3242 & 3.4259 & 0.6583 & 0.3374 & 17.77 & 0.5546 & 0.2643 & 1.9287 & 0.6461 & 0.4133 & 20.26 & 0.5567 & 0.2757 & 3.1648 & 0.6239 & 0.4398 \\ 
    FeMaSR & 19.38 & 0.6044 & 0.3193 & 4.3513 & 0.6486 & 0.3586 & 16.78 & 0.5580 & 0.2274 & 4.2489 & 0.6234 & 0.4429 & 19.28 & 0.5960 & 0.2127 & 3.6849 & 0.5838 & 0.4522 \\ 
    Swin2SR & \redb{20.97} & \redb{0.6577} & 0.4307 & 3.8591 & 0.6103 & 0.3986 & \ublue{20.01} & \ublue{0.6782} & 0.3591 & 4.6336 & 0.5821 & 0.4864 & 22.29 & 0.7138 & 0.2724 & 4.5726 & 0.5663 & 0.4616 \\ \hdashline 
    \textbf{StarSRGAN} & 20.35 & \ublue{0.6496} & \redb{0.2302} & \ublue{2.8900} & \redb{0.7127} & \redb{0.4508} & 19.94 & \redb{0.7167} & \redb{0.1045} & \ublue{1.9064} & \redb{0.7299} & \redb{0.5538} & \redb{24.61} & \redb{0.8065} & 0.1456 & 3.0859 & \redb{0.6943} & \redb{0.5626} \\
    \textbf{StarSRGAN Lite} & 20.36 & 0.5736 & 0.2912 & 3.1389 & 0.6124 & 0.3396 & \redb{20.50} & 0.6736 & 0.2941 & 3.6585 & 0.6035 & 0.4790 & 22.46 & \ublue{0.7195} & 0.1919 & 3.1200 & 0.6128 & 0.4994 \\ \hline
    \end{tabular}
    }
    \label{table:overall}
\end{table*}

\begin{equation}
\begin{split}
l_{V G G / i, j}=&\frac{1}{C_{i,j}W_{i, j} H_{i, j}} \sum_{z=1}^{C_{i, j}} \sum_{x=1}^{W_{i, j}} \sum_{y=1}^{H_{i, j}}\\
&\left|\Phi_{i, j}\left(G_{\theta_{G}}\left(I^{L R}\right)\right)_{x, y, z}-\Phi_{i, j}\left(I^{H R}\right)_{x, y, z}\right|\label{VGG_loss}
\end{split}
\end{equation}
where $\Phi_{i, j}$ represents features acquired by the $j\mbox{-}th$ convolution (before activation) in the VGG network before the $i\mbox{-}th$ max-pooling layer. In the VGG network, $C_{i, j}$, $W_{i, j}$, and $H_{i, j}$ are the dimensions of their respective feature spaces.

Based on the concepts of SRGAN \cite{Led17},  the ResNet loss is defined based on the ReLU activation layer of the 50-layer pre-trained ResNet network presented in \cite{Kai16}. Since the ResNet network structure differs from the VGG network, each feature space is specified by a unique block. ResNet-50 architecture is divided into four stages, each containing several bottleneck layers. The extracted perceptual features use the output value of the bottleneck layer at each stage, and the ResNet loss can also be expressed as:
\begin{equation}
\begin{split}
l_{R E S / m, n}=&\frac{1}{C_{m,n} W_{m, n} H_{m, n}} \sum_{z=1}^{C_{m, n}} \sum_{x=1}^{W_{m, n}} \sum_{y=1}^{H_{m, n}}\\
&\left|\beta_{m, n}\left(G_{\theta_{G}}\left(I^{L R}\right)\right)_{x, y, z}-\beta_{m, n}\left(I^{H R}\right)_{x, y, z}\right|\label{RES_loss}
\end{split}
\end{equation}
where $\beta_{m, n}$ represents features obtained by the $n\mbox{-}th$ bottleneck layer (after activation) at the $m\mbox{-}th$ stage. $C_{m, n}$, $W_{m, n}$, and $H_{m, n}$ are the dimensions of their respective feature spaces in the ResNet network.

Finally, the Dual Perceptual Loss $l_{percep}$ function under the two perceptual losses is expressed as:
\begin{equation} 
l_{D P}=l_{V G G}+\frac{1}{\mu} \zeta_{l_{V G G}, l_{R E S}} l_{R E S}\label{DP_loss}, 
\end{equation}
where the ResNet loss $l_{R E S}$ is weighted dynamically and the weight value is $\frac{1}{\mu} \zeta_{l_{V G G}, l_{R E S}}$, $\mu$ is a nonzero constant. The $\zeta_{l_{V G G}, l_{R E S}}$ can be expressed as:
\begin{equation}
\zeta_{l_{V G G}, l_{R E S}}=\frac{l_{V G G}+c}{l_{R E S}+c}
\label{ratio}
\end{equation}
Where $c$ is a small positive constant when its job only prevents the denominator from becoming zero, therefore, $\frac{1}{\mu} \zeta_{l_{V G G}, l_{R E S}}$  is only a value that fluctuates with the ratio of $l_{V G G}$ to $l_{R E S}$. Consequently, $\frac{1}{\mu} \zeta_{l_{V G G}, l_{R E S}}$ is regarded as the weight value under the ResNet loss, which can only alter the update range of network parameters and not the update direction.

\begin{table*}[t]
    \caption{Upsampling benchmark and inference performance of Real-ESRGAN and StarSRGAN models. \\
    System specification: Ubuntu 22.04 LTS, AMD Ryzen 5 5600X, NVIDIA GeForce RTX 3080 Ti, CUDA 12.1} \label{tab:synthetic}
    \setlength{\tabcolsep}{2pt}
    \renewcommand*{\arraystretch}{1.2}
    \resizebox{\textwidth}{!}{
    \begin{tabular}{|c|cc|cc|cc|cc|cc|cc|cc|}
    \hline
    \multirow{3}{*}{Method} & \multicolumn{10}{c|}{Upsampling Benchmark (FPS)} & \multicolumn{4}{c|}{Performance} \\ \cline{2-15}
     & \multicolumn{2}{c|}{360p to 1080p} & \multicolumn{2}{c|}{480p to 1440p} & \multicolumn{2}{c|}{540p to 4K} & \multicolumn{2}{c|}{720p to 5K} & \multicolumn{2}{c|}{1080p to 8K} & \multicolumn{2}{c|}{Image Quality} & \multicolumn{2}{c|}{Inference Time} \\ \cline{2-15}
     & Python & C++ & Python & C++ & Python & C++ & Python & C++ & Python & C++ & NR-IQA & FR-IQA & PyTorch & NCNN \\ \hline
    Real-ESRGAN & 2.85 & 6.94 & 1.63 & 3.97 & 1.29 & 3.14 & 0.73 & 1.78 & 0.30 & 0.73 & 100\% & 100\% & 100\% & 100\% \\ 
    StarSRGAN & 2.59 & 6.42 & 1.56 & 3.81 & 1.24 & 3.02 & 0.69 & 1.72 & 0.28 & 0.69 & 107\%	& 112\%	& 94\% & 95\% \\
    StarSRGAN Lite & 21.32 & 53.68 & 11.67 & 28.43 & 9.36 & 22.79 & 5.44 & 13.59 & 2.48 & 6.14 & 89\% & 93\% & 739\% & 753\% \\\hline
    \end{tabular}
    }
    \label{table:fps}
\end{table*}

\begin{table*}[t]
    \caption{Quantitative comparison of StarSRGAN variations on common test datasets for SR (average IQA). LPIPS/NIQE $\downarrow$: the lower, the better. PSNR/SSIM/MANIQA/AHIQ $\uparrow$: the higher, the better.  Note that MANIQA and AHIQ are the current SOTA IQA methods. The best and second performance are marked in \redb{red} and \ublue{blue}.\\
    \textbf{Discussion}: The dropout degradation technique needs to be more predictable when implement on SR models. Some models even have better NIQE than the HR. The IQA may not be accurately sufficient for blind SR tasks.} \label{tab:synthetic}
    \setlength{\tabcolsep}{3pt}
    \renewcommand*{\arraystretch}{1.2}
    \resizebox{\textwidth}{!}{
    \begin{tabular}{|c|c|c|c|c|c|cccc|cc|}
    \hline
    Method & Adapt-Deg & Attn-Unet & Multi-Disc & Dual-Loss & Drop-Out & PSNR & SSIM & LPIPS & NIQE & MANIQA & AHIQ \\ \hline
    Original HR & \cellcolor{black!10} & \cellcolor{black!10} & \cellcolor{black!10} & \cellcolor{black!10} & \cellcolor{black!10} & 80.00 & 1.0000 & 0.0000 & 3.4135 & 0.6986 & 0.6543 \\
    Real-ESRGAN & \cellcolor{black!10} & \cellcolor{black!10} & \cellcolor{black!10} & \cellcolor{black!10} & \cellcolor{black!10} & 20.94 & 0.6166 & 0.2618 & 4.1181 & 0.6378 & 0.4818 \\ 
    \hline \hline
    StarSRGAN V1 & \cellcolor{black!10} & \cellcolor{black!10} & \cellcolor{black!10} & \cellcolor{black!10} & \cellcolor{black!10} & 23.43 & 0.6893 & 0.1361 & 2.5619 & 0.6320 & 0.5073 \\
    StarSRGAN V2 & \cellcolor{green}\checkmark & \cellcolor{black!10} & \cellcolor{black!10} & \cellcolor{black!10} & \cellcolor{black!10} & \ublue{24.56} & \ublue{0.7152} & 0.1220 & 2.7378 & 0.6422 & 0.5154 \\
    StarSRGAN V3 & \cellcolor{green}\checkmark & \cellcolor{green}\checkmark & \cellcolor{black!10} & \cellcolor{black!10} & \cellcolor{black!10} & 24.38 & 0.7065 & 0.1226 & 2.7281 & 0.6582 & 0.5442 \\  
    StarSRGAN V4 & \cellcolor{green}\checkmark & \cellcolor{green}\checkmark & \cellcolor{green}\checkmark & \cellcolor{black!10} & \cellcolor{black!10} & 24.43 & \redb{0.7185} & 0.1211 & 2.8669 & \ublue{0.6590} & 0.5333 \\
    StarSRGAN V5 & \cellcolor{green}\checkmark & \cellcolor{green}\checkmark & \cellcolor{green}\checkmark & \cellcolor{green}\checkmark & \cellcolor{black!10} & 23.21 & 0.7117 & 0.1518 & 2.9507 & \redb{0.6608} & 0.5504 \\ 
    \hline \hline
    StarSRGAN+ V1 & \cellcolor{black!10} & \cellcolor{black!10} & \cellcolor{black!10} & \cellcolor{black!10} & \cellcolor{green}\checkmark & 23.12 & 0.6769 & 0.1361 & 2.6444 & 0.6368 & 0.5483 \\
    StarSRGAN+ V2 & \cellcolor{green}\checkmark & \cellcolor{black!10} & \cellcolor{black!10} & \cellcolor{black!10} & \cellcolor{green}\checkmark & 23.86 & 0.7144 & 0.1475 & 2.4271 & 0.6458 & \ublue{0.5681} \\
    StarSRGAN+ V3 & \cellcolor{green}\checkmark & \cellcolor{green}\checkmark & \cellcolor{black!10} & \cellcolor{black!10} & \cellcolor{green}\checkmark & 23.60 & 0.7035 & \redb{0.1187} & \ublue{2.3328} & 0.6421 & \redb{0.5705} \\  
    StarSRGAN+ V4 & \cellcolor{green}\checkmark & \cellcolor{green}\checkmark & \cellcolor{green}\checkmark & \cellcolor{black!10} & \cellcolor{green}\checkmark & 23.60 & 0.7031 & \ublue{0.1206} & 2.3468 & 0.6403 & 0.5673 \\
    StarSRGAN+ V5 & \cellcolor{green}\checkmark & \cellcolor{green}\checkmark & \cellcolor{green}\checkmark & \cellcolor{green}\checkmark & \cellcolor{green}\checkmark & \redb{24.75} & 0.7142 & 0.1255 & \redb{2.2946} & 0.6444 & 0.5357 \\  
    \hline \hline
    StarSRGAN Lite V1 & \cellcolor{black!10} & \cellcolor{black!10} & \cellcolor{black!10} & \cellcolor{black!10} & \cellcolor{black!10} & 19.97 & 0.5452 & 0.3684 & 4.0249 & 0.5317 & 0.3410 \\
    StarSRGAN Lite V2 & \cellcolor{green}\checkmark & \cellcolor{black!10} & \cellcolor{black!10} & \cellcolor{black!10} & \cellcolor{black!10} & \redb{22.49} & \redb{0.6524} & \redb{0.2340} & 3.8435 & 0.5344 & 0.4287 \\
    StarSRGAN Lite V3 & \cellcolor{green}\checkmark & \cellcolor{green}\checkmark & \cellcolor{black!10} & \cellcolor{black!10} & \cellcolor{black!10} & 21.68 & 0.6385 & 0.2373 & \redb{3.4966} & 0.5546 & 0.4489 \\  
    StarSRGAN Lite V4 & \cellcolor{green}\checkmark & \cellcolor{green}\checkmark & \cellcolor{green}\checkmark & \cellcolor{black!10} & \cellcolor{black!10} & 22.08 & 0.6373 & \ublue{0.2350} & \ublue{3.6288} & 0.5578 & \redb{0.4732} \\
    StarSRGAN Lite V5 & \cellcolor{green}\checkmark & \cellcolor{green}\checkmark & \cellcolor{green}\checkmark & \cellcolor{green}\checkmark & \cellcolor{black!10} & \ublue{22.30} & \ublue{0.6490} & 0.2479 & 3.7701 & \ublue{0.5639} & \ublue{0.4605} \\ 
    \hline \hline
    StarSRGAN Lite+ V1 & \cellcolor{black!10} & \cellcolor{black!10} & \cellcolor{black!10} & \cellcolor{black!10} & \cellcolor{green}\checkmark & 18.40 & 0.5688 & 0.2960 & 3.6561 & 0.5161 & 0.3842 \\
    StarSRGAN Lite+ V2 & \cellcolor{green}\checkmark & \cellcolor{black!10} & \cellcolor{black!10} & \cellcolor{black!10} & \cellcolor{green}\checkmark & 20.30 & 0.5609 & 0.2472 & 3.7053 & 0.5635 & 0.4518 \\
    StarSRGAN Lite+ V3 & \cellcolor{green}\checkmark & \cellcolor{green}\checkmark & \cellcolor{black!10} & \cellcolor{black!10} & \cellcolor{green}\checkmark & 18.73 & 0.5693 & 0.2491 & 3.9238 & 0.5614 & 0.4487 \\   
    StarSRGAN Lite+ V4 & \cellcolor{green}\checkmark & \cellcolor{green}\checkmark & \cellcolor{green}\checkmark & \cellcolor{black!10} & \cellcolor{green}\checkmark & 19.65 & 0.5646 & 0.2363 & 4.9272 & \redb{0.5813} & 0.4569 \\
    StarSRGAN Lite+ V5 & \cellcolor{green}\checkmark & \cellcolor{green}\checkmark & \cellcolor{green}\checkmark & \cellcolor{green}\checkmark & \cellcolor{green}\checkmark & 19.93 & 0.5714 & 0.2532 & 5.1208 & 0.5197 & 0.4221 \\ \hline
    \end{tabular}
    }
    \label{table:variation}
\end{table*}

\subsection{Dropout Degradation}

In high-level vision tasks, dropout is intended to reduce the overfitting problem. However, it is rarely used in low-level vision tasks such as image SR. As a traditional regression problem, SR behaves differently for high-level tasks and is sensitive to the dropout process. RDSR \cite{Kon21} dropout research demonstrates that appropriate dropout utilization benefits SR networks and improves generalizability. In our study, we employ this approach primarily for observational purposes.

In particular, we add the dropout layer before the final output layer. According to the results of our experiments, this technique improves network performance in a multi-degradation condition.

\section{Experiments}

\subsection{Implementation}

\begin{figure*}
	\vspace{-0.2cm}
	\begin{center}
		\includegraphics[width=1\linewidth]{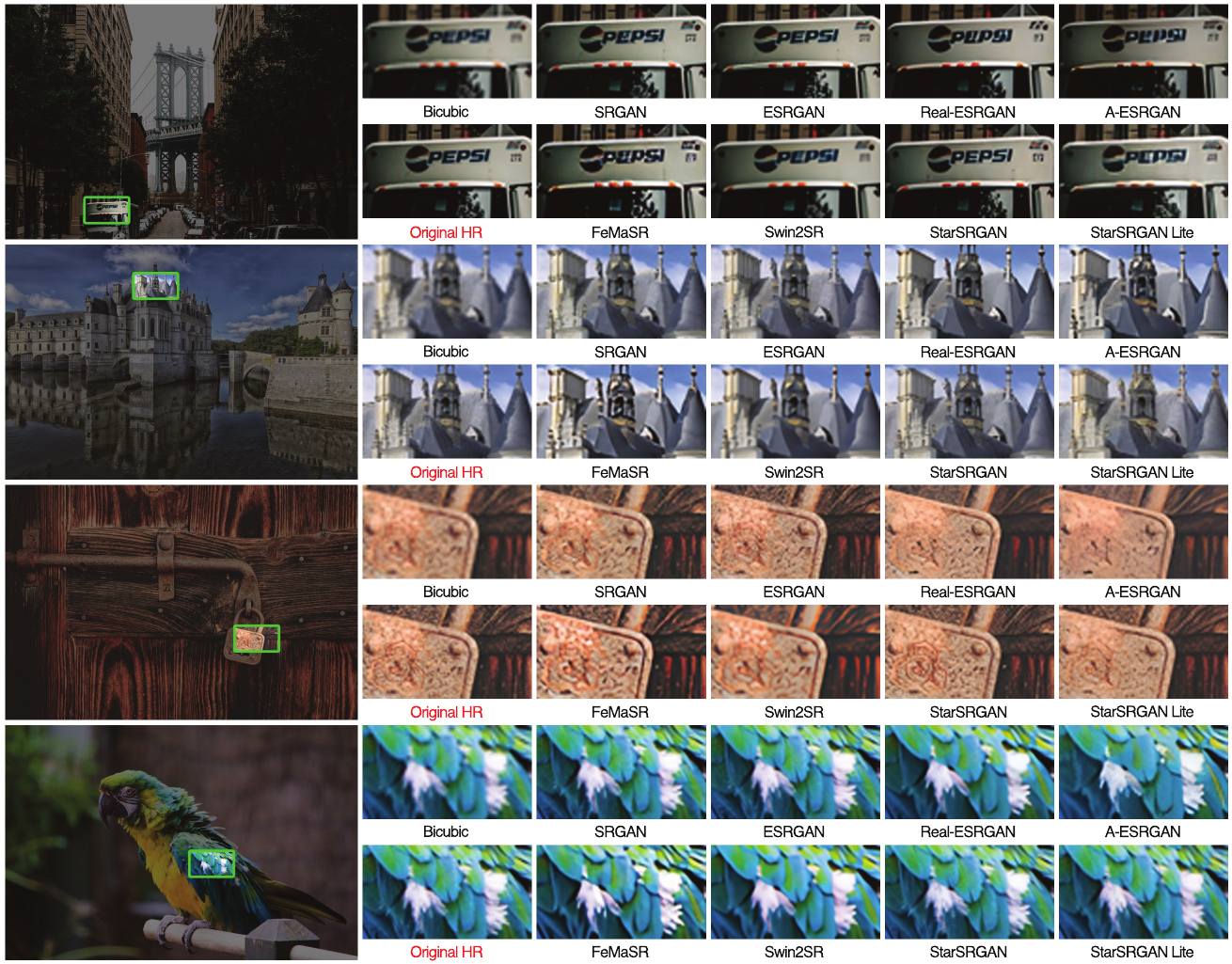}
	\end{center}
	\caption{Visual comparisons of standard SR models with StarSRGAN models in $4\times$. (\textbf{Zoom in for best view})\\
            \textbf{Discussion}: The logo on the truck and the carving of the lock were recovered precisely on StarSRGAN. The image quality reconstructed by StarSRGAN Lite stays caught up with other models. All models deliver pleasant details and textures on the parrot feathers and the bell tower. Artifacts are also hard to be spotted on every model.}
	\label{fig:v4}
\end{figure*}

To better compare the functionality of various mechanisms, including:
Adaptive Degradation Model (Adapt-Deg), Attention U-Net Discriminator (Attn-Unet), Multi-scale Discriminator (Multi-Disc), and Dual Perceptual Loss (Dual-Loss), we build 5 different StarSRGAN models corresponding models, including:
\begin{itemize}
\item StarSRGAN (V1): nearest similar to Real-ESRGAN but using the novel Star Dense Block.
\item StarSRGAN V1 + Adapt-Deg (V2): using the Apdative Degradation Model instead of the High-order Degradation Model.
\item StarSRGAN V2 + Attn-Unet (V3): using the Attention U-Net Discriminator instead of U-Net Discriminator.
\item StarSRGAN V3 + Multi-Disc (V4): using the Multi-scale Discriminator instead of Single Discriminator.
\item StarSRGAN V4 + Dual-Loss (V5): using the Dual Perceptual Loss instead of Single Perceptual Loss.
\end{itemize}
Similarly, we also have 5 StarSRGAN Lite models (from V1 to V5), representing corresponding variations of StarSRGAN in compact architecture.

To observe the benefit of the dropout degradation (Drop-Out) technique, we also conducted a separate similar experiment with dropout layers, denoted by the plus sign (e.g. StarSRGAN+, StarSRGAN Lite+).

We trained 5 StarSRGAN, 5 StarSRGAN+, 5 StarSRGAN Lite, and 5 StarSRGAN Lite+ models on DIV2K \cite{Nti17a} and Flickr2K \cite{Nti17b} datasets. The training HR size is set to 256. We train our models on an NVIDIA A100 GPU with a batch size of 32 by using Adam optimizer. We train the StarSRNet models for $2000K$ iterations with a learning rate $2\times 10^{-4}$ while training the StarSRGAN models for $1000K$ iterations with a learning rate $1\times 10^{-4}$. We also adopt the Exponential Moving Average (EMA) for more regular training and better performance. StarSRGAN and StarSRGAN Lite are trained with a combination of L1 Loss, Perceptual Loss, and GAN loss, with weights $[1, 1, 0.1]$, respectively. Models which implement the Multi-scale Discriminator are composed of two discriminators, $D_{normal}$ and $D_{sampled}$, which have the input of $1 \times$ and $2 \times$ down-sampled images as the input. The weight for GAN loss of $D_{normal}$ and $D_{sampled}$ is $[1, 1]$. Our implementation is based on the BasicSR \cite{Wan22}.

\begin{figure*}
	\vspace{-0.2cm}
	\begin{center}
		\includegraphics[width=1\linewidth]{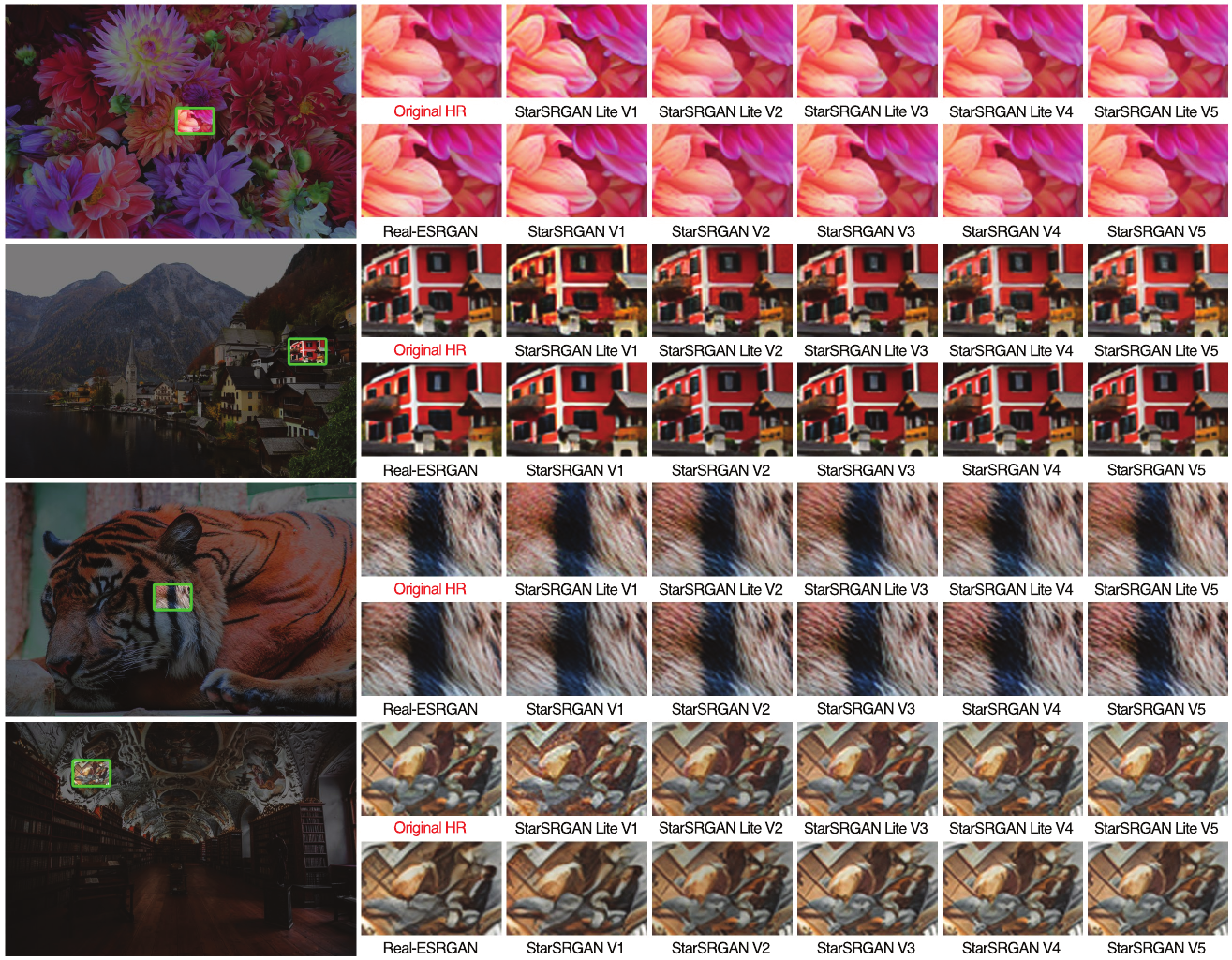}
	\end{center}
	\caption{Visual comparisons of StarSRGAN variations with an upsampling scale of $4\times$. (\textbf{Zoom in for best view})\\
 \textbf{Discussion}: If we pay close attention, we can observe that the StarSRGAN V5 delivered more tiny details on the petals. The chimneys and the windows also look more similar to the HR. Every model reconstructed the fur of the tiger with outstanding quality. The coat color was also better restored on StarSRGAN V5 than on StarSRGAN V1 or Real-ESRGAN. The images enhanced by StarSRGAN Lite models are acceptable compared to its predecessor.}
	\label{fig:star}
\end{figure*}

\subsection{Datasets}

Previous studies have typically evaluated blind image SR models using synthetic LR images manually degraded from HR images. However, these images may not accurately represent the LR images resulting from real-world degradation processes, which often involve complex combinations of multiple degradation processes. Additionally, no publicly available datasets contain LR images from real-world sources. As an alternative, we have used real-world images scaled up by a factor of 4 for testing purposes.
We employ real-world images from 5 classical benchmarks to evaluate our approach, including Set5, Set14, BSD100, Urban100, Manga109, and the modern DIV2K Validation dataset \cite{Nti17a}. These datasets contain images from diverse categories, such as portraits, landscapes, and structures. A reliable SR model should perform well on most of these standard datasets.

\subsection{Compared Methods}

\begin{figure*}
	\vspace{-0.2cm}
	\begin{center}
		\includegraphics[width=1\linewidth]{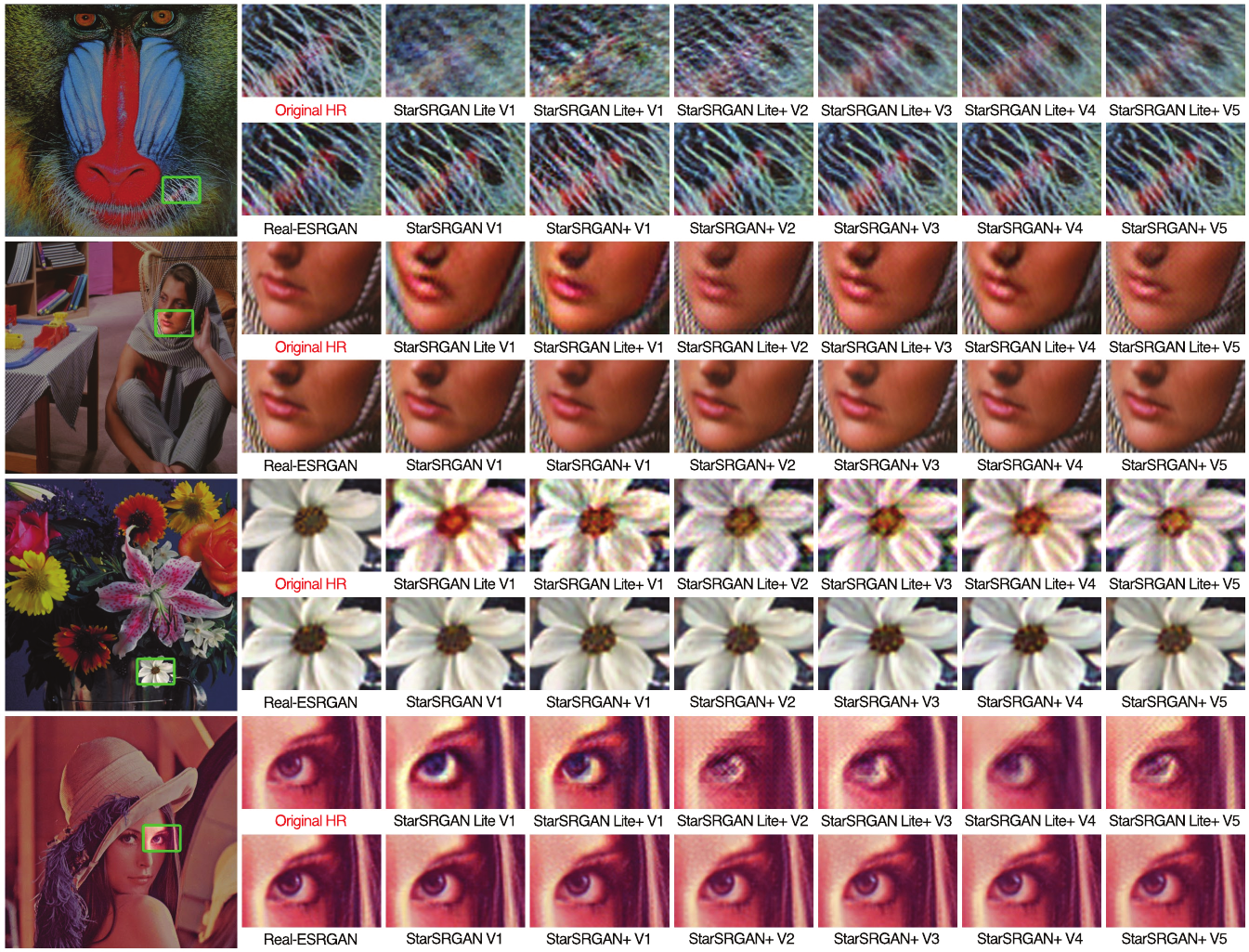}
	\end{center}
	\caption{Further observation of StarSRGAN+ models applying Dropout technique. (\textbf{Zoom in for best view})\\
     \textbf{Discussion}: We can easily observe that Dropout Degradation brings unstable performance on StarSRGAN Lite models. On the contrary, the technique adequately integrated with StarSRGAN without apparent conflicts.}
	\label{fig:starplus}
\end{figure*}

We examine the proposed StarSRGAN and StarSRGAN Lite models with the SRGAN \cite{Led17}, ESRGAN \cite{Wan19}, Real-ESRGAN \cite{Wan21}, A-ESRGAN \cite{Wei21}, FeMaSR \cite{Che22}, and Swin2SR \cite{Con22} models. The architecture of StarSRGAN V1  is the nearest similar to the architecture of Real-ESRGAN. More specifically, Residual Dense Block has been replaced with the novel Star Residual Dense Block, which can help evaluate the effectiveness of StarSRGAN even with a slight modification. On the other hand, StarSRGAN Lite models aim to reduce the reconstruction time of super-resolution. Therefore, in addition to comparing perceptual quality, we also compare reconstruction time in frame rate (expressed in frames per second or FPS) between the Real-ESRGAN and StarSRGAN, and StarSRGAN Lite models.

\subsection{Experiment Results}

Table \ref{table:overall} compares StarSRGAN with other SR models on several standard test datasets for SR. The results indicate that the classical IQAs like PSNR, SSIM, and NIQE are no more suitable for SR evaluation. Previous research claims that to achieve high PNSR, the model tends to generate over-smooth results. The baseline model, SRGAN, usually obtains a better SSIM index than more advanced models. Moreover, some models even have a better NIQE score than the HR, making this measure the most unpredictable IQA in this work. On the other hand, the measurements that come from MANIQA and AHIQ are anticipated and reasonable.

Models with real-world emulated data synthesis perform poorly on illustration. Fortunately, StarSRGAN with Adaptive Degradation Model has partially solved this issue. Unnatural city-featured textures like windows, roads, and bricks tend to be more complicated to reconstruct than other textures. Fine-tuning models on extra training data is promising to sort out the problem.

Through upsampling benchmark results shown in  Table \ref{table:fps}, we can find that StarSRGAN has traded off its inference time to achieve better image quality compared to Real-ESRGAN. In the opposite direction, StarSRGAN Lite sacrifices its image quality to gain impressive performance in reconstructed time. Specifically, the lightweight architecture has brought real-time performance with more than 20 FPS when upscaling from 540p to 4K with C++ optimized executable file.

Table \ref{table:variation} shows that even our most straightforward variation, StarSRGAN V1, outperforms the Real-ESRGAN method in most metrics. Variations applied Dropout Degradation technique are not steady enough, and further research should be conducted. From visual comparison (some examples are shown in Figure \ref{fig:star} and Figure \ref{fig:starplus}), we observe that our methods can recover sharper edges and restore better texture details. Although, it is hard for human-eye to distinguish because both Real-ESRGAN and StarSRGAN bring excellent perceptual quality. Note that the StarSRGAN V5 models have been selected for comparison in Table \ref{table:overall} and Table \ref{table:fps}.

\section{Conclusions}

The present study introduces two novel GAN-based models, StarSRGAN and StarSRGAN Lite, for blind SR tasks. StarSRGAN integrates advancements from 5 previous research works and yields new SOTA performance, surpassing the leading SR method, Real-ESRGAN, by 10\% on both SOTA No-Reference IQA and Full-Reference IQA methods (MANIQA and AHIQ). StarSRGAN Lite, a lightweight version of StarSRGAN, also inherits improvements from its predecessor and offers real-time inference performance, processing upsampled frames from 540p to 4k at over 20 FPS when executed on a C++ optimized executable file. Several directions for further enhancement of StarSRGAN architectures are recommended. For instance, retraining the models with newly released datasets such as DIV8K or Unsplash. Additionally, other activation functions such as SiLU and GELU, could be a better alternative for the familiar ReLU. Super-resolving only interested objects and disregarding unnecessary regions like the background could improve StarSRGAN inference performance. Applying Video SR techniques and leveraging spatiotemporal data could also be a promising direction for further research. Currently, StarSRGAN models support only the $4\times$ upscale factor, and other upscale factors such as $2\times$, $8\times$, and $16\times$ are also necessary. Another approach is employing an image classifier to distinguish between real-life and unreal images and choosing the most optimized for each case. Better batch size and more iterations could be explored with more robust hardware. In conclusion, these directions could facilitate the development of future research.

\textbf{Acknowledgement.}  This research is funded by University of Science, VNU-HCM project CNTT 2023-08.

%
%


\end{document}